\documentclass[conference]{IEEEtran}
\IEEEoverridecommandlockouts

\usepackage{cite}
\usepackage{amsmath,amssymb,amsfonts,amsthm}
\usepackage[pdftex]{graphicx}
\usepackage{textcomp}
\usepackage{xcolor}
\usepackage{booktabs}
\usepackage{multirow}
\usepackage{multicol}
\usepackage{graphicx}
\usepackage[T1]{fontenc}
\usepackage[utf8]{inputenc}
\usepackage{babel}
\usepackage{mathtools}
\usepackage{algorithm}
\usepackage{tikz}
\usepackage{pgfplots}
\usepackage{glossaries-extra}
\usepackage{siunitx}
\usepackage{algpseudocodex}
\usepackage{svg}
\usepackage{etoolbox}
\usepackage{balance}
\usepackage{microtype}
\usepackage{mathtools}
\usepackage[hidelinks]{hyperref}
\usepackage[capitalise]{cleveref}
\usepackage{xcolor, soul}
\sethlcolor{red}

\setabbreviationstyle[acronym]{long-short}

\svgsetup{inkscapelatex=false}

\pgfplotsset{compat=newest,
every axis/.style={
    grid=major,
    xlabel near ticks,
    ylabel near ticks,
    legend pos=south east,
    legend style={font=\footnotesize},
    every x tick label/.append style = {font=\footnotesize},
    every y tick label/.append style = {font=\footnotesize},
    cycle list/Set2-8,
    cycle multiindex* list={
    mark list\nextlist
    Set2-8\nextlist
    linestyles\nextlist
    },
    every axis plot/.append style={mark size=1pt, thick},
    enlarge y limits=0.025,
    enlarge x limits=0
}
}
\graphicspath{{figures/}}

\usepgfplotslibrary{colormaps}
\usepgfplotslibrary{colorbrewer}
\usepgfplotslibrary{external}

\makeatletter
\newcommand\fs@betterruled{%
  \def\@fs@cfont{\bfseries}\let\@fs@capt\floatc@ruled
  \def\@fs@pre{\vspace*{5pt}\hrule height.8pt depth0pt \kern2pt}%
  \def\@fs@post{\kern2pt\hrule\relax}%
  \def\@fs@mid{\kern2pt\hrule\kern2pt}%
  \let\@fs@iftopcapt\iftrue}
\floatstyle{betterruled}
\restylefloat{algorithm}
\makeatother

\usepackage{tikz}
\usetikzlibrary{backgrounds, calc, chains, fit, matrix, positioning, shadows, shapes, intersections, circuits.ee}
\tikzset{input/.style={}}
\tikzset{output/.style={}}
\tikzset{op/.style={circle, draw, fill=black!10, minimum size=2.5ex, inner sep=0ex}}
\tikzset{filter/.style={rectangle, draw, thick, fill=black!10, minimum size=3.5ex, inner sep=1ex}}
\tikzset{nn/.style={trapezium, trapezium angle=80, draw, thick, fill=black!10, inner sep=1ex}}
\tikzset{branch/.style={circle, draw, thick, fill=black, minimum size=.5ex, inner sep=0ex}}
\tikzset{tensor/.style={rectangle, draw, fill=white, minimum size=2em, double copy shadow={shadow xshift=.5ex,shadow yshift=-.5ex}}}
\tikzset{rounded/.style={rounded rectangle, draw, thick, fill=black!10, minimum size=3.5ex, inner xsep=1ex}}
\tikzset{image/.style={rectangle, draw, fill=white, minimum size=2em}}
\tikzset{>=direction ee}
\tikzset{/tikz/thin/.style={line width=.9pt}}
\tikzset{/tikz/thick/.style={line width=1.4pt}}
\tikzset{every path/.style={thin}}

\usepackage[none]{hyphenat}
\usepackage{subfigure}

\renewcommand{\vec}[1]{\mathbf{#1}}
\newcommand{\vecs}[1]{\boldsymbol{#1}}

\newcommand{\nv}{\vec{n}}

\newcommand{\xv}{\vec{x}}
\newcommand{\yv}{\vec{y}}
\newcommand{\zv}{\vec{z}}

\newcommand{\psiv}{\vecs{\psi}}

\newcommand{\Am}{\vec{A}}

\newcommand{\Thetav}{\vecs{\Theta}}
\newcommand{\Phiv}{\vecs{\Phi}}

\newcommand{\Cc}{{\cal C}}

\newcommand{\Lc}{{\cal L}}

\newcommand{\Nc}{{\cal N}}

\newcommand{\Pc}{{\cal P}}

\newcommand{\CC}{\mathbb{C}}

\newcommand{\NN}{\mathbb{N}}
\newcommand{\RR}{\mathbb{R}}
\newcommand{\II}{\mathbb{I}}

\newcommand{\LB}{\left(}
\newcommand{\RB}{\right)}

\newcommand{\LSB}{\left[}
\newcommand{\RSB}{\right]}

\newcommand{\EE}{{\mathbb{E}}}

\newcommand{\argmin}[1]{\underset{#1}{\operatorname{arg}\,\operatorname{min}}\;}

\newcommand\norm[1]{\left\lVert#1\right\rVert}

\newcommand{\logn}[2]{\mathop{\mathrm{log}_{#1} \LB #2\RB}}

\newcommand{\Pavg}{P_\mathrm{avg}}

\newcommand{\removed}[1]{}
\newcommand{\selim}[1]{#1} 

\newacronym{ACM}{ACM}{adaptive coding and modulation}
\newacronym{ADC}{ADC}{analog-to-digital conversion}
\newacronym{AGC}{AGC}{automatic gain control}
\newacronym{AWGN}{AWGN}{additive white Gaussian noise}
\newacronym{BER}{BER}{bit error rate}
\newacronym{BLER}{BLER}{block error rate}
\newacronym{BP}{BP}{backpropagation}
\newacronym{BPTT}{BPTT}{backpropagation through time}
\newacronym{CE}{CE}{cross-entropy}
\newacronym{CFO}{CFO}{carrier frequency offset}
\newacronym{CSI}{CSI}{channel state information}
\newacronym{DAC}{DAC}{digital-to-analog conversion}
\newacronym{DL}{DL}{deep learning}
\newacronym{DFT}{DFT}{discrete Fourier transform}
\newacronym{FFT}{FFT}{fast Fourier transform}
\newacronym{GAN}{GAN}{generative adversarial network}
\newacronym{GRU}{GRU}{gated recurrent unit}
\newacronym{iid}{i.i.d.\@}{independent and identically distributed}
\newacronym{IFFT}{IFFT}{inverse fast Fourier transform}
\newacronym{KL}{KL}{Kullback-Leibler}
\newacronym{LSTM}{LSTM}{long short-term memory}
\newacronym{MDP}{MDP}{Markov decision process}
\newacronym{ML}{ML}{machine learning}
\newacronym{MLP}{MLP}{multilayer perceptron}
\newacronym{MIMO}{MIMO}{multiple-input multiple-output}
\newacronym{MSE}{MSE}{mean squared error}
\newacronym{NN}{NN}{neural network}
\newacronym{DNN}{DNN}{deep neural network}
\newacronym{OFDM}{OFDM}{orthogonal frequency-division multiplexing}
\newacronym{pdf}{pdf}{probability density function}
\newacronym{pmf}{pmf}{probability mass function}
\newacronym{PSNR}{PSNR}{peak signal to noise ratio}
\newacronym{RBF}{RBF}{Rayleigh block-fading}
\newacronym{ReLU}{ReLU}{rectified linear unit}
\newacronym{RL}{RL}{reinforcement learning}
\newacronym{RNN}{RNN}{recurrent neural network}
\newacronym{SFO}{SFO}{sampling frequency offset}
\newacronym{SNR}{SNR}{signal-to-noise ratio}
\newacronym{SINR}{SINR}{signal-to-interference-plus-noise ratio}
\newacronym{SGD}{SGD}{stochastic gradient descent}
\newacronym{wrt}{w.r.t.\@}{with respect to}

\newacronym{OAC}{OAC}{over-the-air computation}
\newacronym{MAC}{MAC}{multiple access channel}
\newacronym{SIC}{SIC}{successive interference cancellation}
\newacronym{TDMA}{TDMA}{time division multiple access}
\newacronym{NOMA}{NOMA}{non-orthogonal multiple access}
\newacronym{CL}{CL}{curriculum learning}
\newacronym{JSCC}{JSCC}{joint source-channel coding}
\newacronym{DeepJSCC}{DeepJSCC}{deep joint source-channel coding}
\newacronym{MTL}{MTL}{multi-task learning}
\newacronym{MIL}{MIL}{multi-instance learning}
\newacronym{DML}{DML}{deep metric learning}
\newacronym{IoT}{IoT}{Internet of Things}
\newacronym{SSIM}{SSIM}{structural similarity index measure}
\newacronym{MS-SSIM}{MS-SSIM}{multi-scale \gls{SSIM}}
\newacronym{DDPM}{DDPM}{denoising diffusion probabilistic model}
\newacronym{MVL}{MVL}{multi-view learning}
\newacronym{CNN}{CNN}{convolutional neural network}
\newacronym{LPIPS}{LPIPS}{learned perceptual image patch similarity}
\newacronym{BPG}{BPG}{Better Portable Graphics}

\newacronym{IoE}{IoE}{Internet of Everything}

\newacronym{V2X}{V2X}{vehicle-to-everything}

\newacronym{AR/VR}{AR/VR}{augmented/virtual reality}

\newacronym{DSC}{DSC}{distributed source coding}

\newacronym{ANN}{ANN}{artificial neural network}

\newacronym{BCR}{BCR}{bandwidth compression ratio}
\newacronym{BR}{BR}{bandwidth ratio}
\newacronym{OPTA}{OPTA}{Optimal Performance Theoretically Attainable}
\newacronym{AF}{AF}{attention feature}

\newacronym{DDNM}{DDNM}{deep denoising null-space model}

\newacronym{IR}{IR}{image restoration}
\newacronym{GT}{GT}{ground-truth}
\newacronym{SVD}{SVD}{singular value decomposition}
\title{High Perceptual Quality Wireless Image Delivery \\ with Denoising Diffusion Models}
\title{High Perceptual Quality Wireless Image Delivery \\ with Denoising Diffusion Models
\thanks{The present work has received funding from the European Union’s Horizon 2020 Marie Skłodowska Curie Innovative Training Network Greenedge (GA. No. 953775). \selim{For the purpose of open access, the authors have applied a Creative Commons Attribution (CC BY) license to any Author Accepted Manuscript version arising from this submission.}}
}
\author{\IEEEauthorblockN{Selim F. Yilmaz$^{\star}$, Xueyan Niu$^{\dagger}$, Bo Bai$^{\dagger}$, Wei Han$^{\dagger}$, Lei Deng$^{\dagger}$ and Deniz Gündüz$^{\star}$}
\IEEEauthorblockA{$^{\star}$Imperial College London, London, UK, \{s.yilmaz21,d.gunduz\}@imperial.ac.uk \\ $^{\dagger}$Huawei Technologies Co. Ltd., \{niuxueyan3, baibo8, harvey.hanwei, deng.lei2\}@huawei.com }
}
\begin{document}
%
\maketitle

\begin{abstract}
We consider the image transmission problem over a noisy wireless channel via deep learning-based joint source-channel coding (DeepJSCC) along with a denoising diffusion probabilistic model (DDPM) at the receiver. Specifically, we are interested in the perception-distortion trade-off in the practical finite block length regime, in which separate source and channel coding can be highly suboptimal. We introduce a novel scheme, where the conventional DeepJSCC encoder targets transmitting a lower resolution version of the image, which later can be refined thanks to the generative model available at the receiver. In particular, we utilize the range-null space decomposition of the target image; DeepJSCC transmits the range-space of the image, while DDPM progressively refines its null space contents. Through extensive experiments, we demonstrate significant improvements in distortion and perceptual quality of reconstructed images compared to standard DeepJSCC and the state-of-the-art generative learning-based method. 
\end{abstract}

\section{Introduction}

Traditional wireless communication systems have two essential components: source coding and channel coding. Source coding compresses signals by removing data redundancies, while channel coding introduces structured redundancy into the transmitted signal to improve its resilience against channel noise. 
For wireless image transmission, compression codecs like JPEG or BPG are used to reduce communication resources, but this also lowers the reconstructed image quality. To ensure reliable transmission over a noisy channel, channel coding methods like LDPC, turbo codes, or polar coding are applied. Shannon proved the optimality of this separation-based design in the asymptotic infinite blocklength regime~\cite{shannon1948mathematical}. However, optimality of separation does not hold in real-world scenarios with finite blocklengths. Moreover, suboptimality gap between separation-based and joint schemes enlarges as the blocklength or the channel \gls{SNR} decreases, and such conditions are becoming increasingly more relevant for time-sensitive \gls{IoT} applications~\cite{doudou2012survey}.

\begin{figure*}[tbp!]
    \centering
    \includegraphics[width=\textwidth]{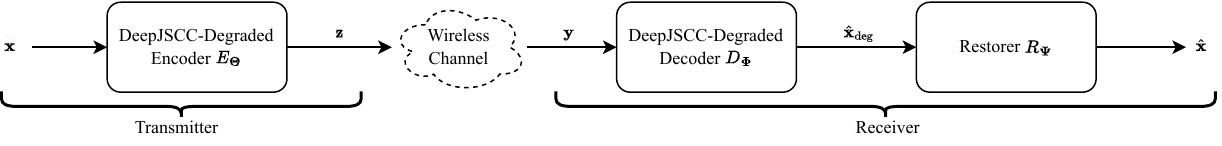}
    \caption{Overview of the image transmission procedure using our method.}
    \label{fig:main_figure}
\end{figure*}

Alternative \gls{JSCC} schemes have long been studied in the literature; however, they have not found applications in practice due to their high complexity and/or limited performance gains with real sources and channels~\cite{ramchandran1993multiresolution, zhai2005joint, bursalioglu2013joint}. 
Recently, there is renewed interest in \gls{JSCC} due to the use of \glspl{DNN}, called \gls{DeepJSCC}~\cite{bourtsoulatze2019deep}. This method models the communication system as a data-driven autoencoder architecture. Follow-up studies on \gls{DeepJSCC} have shown that it can exploit feedback, improve performance by increasing the filter size, and adapt to varying channel bandwidth and \gls{SNR} conditions with almost no loss in performance~\cite{kurka2020deepjscc, burth2020joint, kurka2021bandwidth, xu2021wireless, Yang:TCCN:22, Wu:WCL:22}. It is important to note that \gls{DeepJSCC} avoids the \textit{cliff effect} and achieves \textit{graceful degradation}, which means that the image can still be decoded even if the channel quality falls below the training \gls{SNR}, although with lower reconstruction quality. This provides a major advantage of DeepJSCC compared to separation-based alternatives when accurate channel modelling and estimation is challenging \cite{Anjum:WUWNet:22, Wu:arxiv:23}. A diffusion-based hybrid \gls{DeepJSCC} scheme is also considered in \cite{Niu:SPAWC:23}.


Standard \gls{DeepJSCC} for images~\cite{bourtsoulatze2019deep} focused only on the distortion of the reconstructed image with respect to (w.r.t.) the input. On the other hand, there has been significant progress in recent years in generative models that generate realistic images with better perception qualities. In \cite{Yang:TCCN:22}, adversarial loss has been used for \gls{DeepJSCC} to improve the perceptual quality of the reconstructed images; however, this has achieved limited success due to the difficulties in training, such as maintaining a balance between the loss components and ensuring stable training dynamics. Alternatively, the authors of \cite{erdemir2022generative} employed StyleGAN2, a powerful pretrained generative model to improve the perceptual quality of the image reconstructed by DeepJSCC by modelling the whole encoder-channel-decoder pipeline as a forward process, and modelling the image reconstruction as an inverse problem. With this approach, it is possible to deploy a \gls{DeepJSCC} encoder/decoder pair trained over a generic image dataset, and refine its output by employing a generative model solely at the receiver. In this paper, we follow a similar approach, but use a diffusion-based generative model at the decoder. Our method outperforms both standard \gls{DeepJSCC} and \gls{GAN}-based DeepJSCC of~\cite{erdemir2022generative} on both perception-oriented and distortion-oriented metrics. 


Our main contributions are summarized as follows:

\begin{enumerate}
\item We introduce the first \gls{DDPM}-based \gls{DeepJSCC} scheme for wireless image delivery that utilizes a novel controlled degradation and restoration-based formulation.
\item Through an extensive set of experiments, we demonstrate that our method outperforms conventional \gls{DeepJSCC} and previous state-of-the-art generative learning based \gls{DeepJSCC} for all evaluated \gls{SNR} and bandwidth conditions.
\item To facilitate further research and reproducibility, we provide the source code of our framework and simulations on github.com/ipc-lab/deepjscc-diffusion.
\end{enumerate}


\section{Problem Definition and Methodology}
\label{sec:methodology}
Here, we describe the problem and our novel methodology that combines diffusion models with modified \gls{DeepJSCC}. We decompose our problem into two stages: (1) autoencoding stage, and (2) restoration stage, which are summarized in~\Cref{fig:main_figure}. 

\selim{
\subsection{Notation}
Unless stated otherwise; boldface lowercase letters denote tensors (e.g., $\vec{p}$), non-boldface letters denote scalars (e.g., $p$ or $P$), and uppercase calligraphic letters denote sets (e.g., $\Pc$). $\RR$, $\NN$, $\CC$ denote the set of real, natural and complex numbers, respectively. We define $[n]\triangleq\{1,2,\ldots,n\}$, where $n\in\NN^+$, and $\II \triangleq [255]$.}



\subsection{System Model}
\label{sec:system_model}
We consider wireless image transmission over an \gls{AWGN} channel with noise variance $\sigma^2$. The transmitter maps an input image $\xv \in \II^{C_\mathrm{in} \times W \times H}$, where $W$ and $H$ denote the width and height of the image, while $C_\mathrm{in}$ represents the R, G and B channels for colored images, with a non-linear encoding function $E_{\Thetav}:\LB \II^{C_\mathrm{in} \times W \times H}, \RR \RB \rightarrow \CC^{k}$ parameterized by $\Thetav$ into a complex-valued latent vector $\Tilde{\zv}=E_{\Thetav} (\xv,\sigma)$, where $k$ is the available channel bandwidth. We enforce average transmission power constraint $\Pavg$ on the transmitted signal $\zv \in \CC^k$:
\begin{align}
\label{eq:power_constraint}
\frac{1}{k} \norm{\zv}_2^2 \leq \Pavg.
\end{align}
We satisfy the power constraint by normalizing the signal at the encoder output $\Tilde{\zv}$ via $\zv = \sqrt{k \Pavg/\norm{\tilde{\zv}}_2^2} \tilde{\zv}$.
Then, the transmitter transmits $\zv$ over the \gls{AWGN} channel. The received noisy latent vector is given by $\yv \in \CC^k$ as $\yv = \zv + \nv$, where $\nv \in \CC^{k}$ is \gls{iid} complex Gaussian noise vector with variance $\sigma^2$, i.e., $\nv \sim \Cc\Nc(\vec{0}, \sigma^2 \vec{I}_{k})$. We assume $\sigma$ is known at the transmitters and the receiver.

A non-linear decoding function  $D_{\Phiv}:\LB\CC^{k},\RR\RB\rightarrow \II^{C_{\mathrm{in}}^{'} \times W^{'} \times H^{'}}$ at the receiver, parameterized by $\Phiv$, reconstructs the original image using the channel output $\yv$, i.e.,
$\hat{\vec{x}}_\mathrm{deg} = D_{\Phiv} (\yv, \sigma)$. Lastly, the restorer $R_{\psiv}$ restores the image via $\hat{\xv} = R_{\psiv} \LB \hat{\xv}_\mathrm{deg} \RB$.
Thus, given the channel output $\yv$, the flow of data at the receiver becomes $
\yv \xrightarrow{D_{\Thetav}} \hat{\xv}_\mathrm{deg} \xrightarrow{R_{\psiv}} \hat{\xv}$. Note that, in general, these two steps can be combined into a single decoding step. However, similarly to \cite{erdemir2022generative}, we are interested in refining the degraded image reconstructed by a generic \gls{DeepJSCC} decoder modeled by $D_{\Phiv}$ to improve its perceptual quality. 

The bandwidth ratio $\rho$ characterizes the available channel resources, which is defined as:
\begin{equation*}
    \rho = \frac{k}{ C_\mathrm{in} W H} \;	\;	\;	 \si{channel \;	\;\;		symbols \per pixel}.
\end{equation*}
We also define the $\mathrm{SNR}$, which characterizes the channel quality, as:
\begin{align}
    \label{eq:snr}
    \mathrm{SNR} = 10 \logn{10}{\frac{\Pavg}{\sigma^2}} \,\, \si{\decibel}.
\end{align}
Given $\rho$ and $\mathrm{SNR}$, the goal in general is to minimize the average distortion between the original image $\xv$ at the transmitter and the reconstructed image $\hat{\xv}$ at the receiver, i.e.,
\begin{align}
    \argmin{\Thetav,\Phiv,\psiv} \EE_{r(\xv, \hat{\xv})} \left [ d (\xv, \hat{\xv}) \right ],
\end{align}
where $r(\xv, \hat{\xv})$ is the joint distribution of original image and the reconstructed image and $d (\xv, \hat{\xv})$ can be any distortion metric.
As mentioned, we employ DeepJSCC as the code block, which has been trained to maximize the average \gls{PSNR}, defined as:
\begin{align}
\label{eq:psnr}
d_\mathrm{PSNR} \LB \xv, \hat{\xv} \RB = 10 \logn{10}{\frac{A^2}{
    \frac{1}{C_\mathrm{in}HW}
    \norm{\xv - \hat{\xv}}_2^2 }} \,\, \si{\decibel},
\end{align}
where $A$ is the maximum possible input value, e.g., $A=255$ for images with 8-bit per channel as in our case. 
It is known that \gls{PSNR} is not aligned with human perception in general. Hence, we will also consider the \gls{LPIPS} metric \cite{zhang2018unreasonable}, which has been shown to better match human perception.


\begin{figure*}[tbp!]
    \centering
    \includegraphics[width=\textwidth]{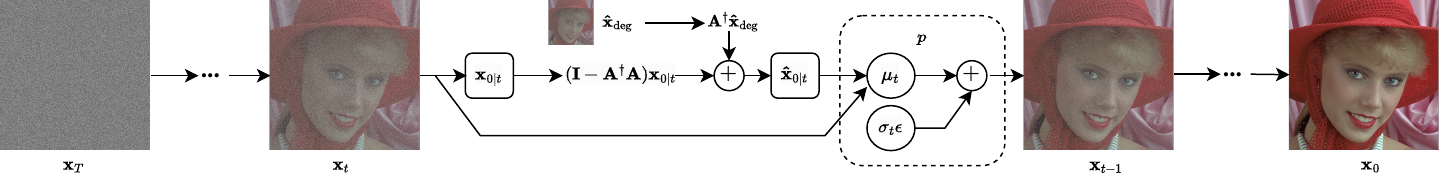}
    \caption{Overview of the image restoration procedure by $R_{\psiv}$.}
    \label{fig:restoration}
\end{figure*}

\begin{algorithm}[tbp!]
\caption{Our \gls{JSCC}-based image transmission procedure}
    \label{alg:overview}
    \begin{algorithmic}
    \LComment{At the transmitter}
    \State $\tilde{\zv} = E_{\Thetav} \LB \xv, \sigma \RB$ \Comment{Encoding}
    \State $\zv = \sqrt{k \Pavg/\norm{\tilde{\zv}}_2^2} \tilde{\zv}$ \Comment{Power normalization}
    \LComment{\Gls{AWGN} channel}
    \State $\yv = \zv + \nv$
    \Comment{Received signal}
    \LComment{At the receiver}
    \State $\hat{\xv}_\mathrm{deg} = D_{\Phiv} (\yv, \sigma)$
    \Comment{Reconstruction of the degraded image}
    \State $\hat{\xv} = R_{\psiv} (\hat{\xv}_\mathrm{deg})$ \Comment{Restoration of the degraded image}
    \State \textbf{return} $\hat{\xv}$
    \end{algorithmic}
\end{algorithm}

\subsection{Transmission of the Degraded Image}


We treat the process from the input image to the reconstructed degraded image as a forward process,  \selim{and we formulate the restoration as an inverse problem.} We would like to approximate the impact of \selim{the} forward process as a known linear transform \selim{that we can invert}, $\Am \in \RR^{d \times D}$.\footnote{We consider flattened versions of $\xv$, $\xv_\mathrm{deg}$, $\hat{\xv}$ and $\hat{\xv}$ for simplicity of the notation, e.g., we map $\xv \in \RR^{C_\mathrm{in} \times W \times H}$ to $\xv \in \RR^{D \times 1}$, where $D=C_\mathrm{in} W H$.} Hence, the aim of DeepJSCC-Degraded is to transmit a degraded version of $\xv$, denoted by $\xv_\mathrm{deg} \in \RR^{d \times 1}$, where
\begin{align*}
    \xv_\mathrm{deg} = \Am \xv. 
\end{align*}
Degradation matrix $\Am$ can be any linear operator, such as decolorization or mean-pooling-based downsampling~\cite{wang2022zero}. Ideal $\Am$ should have the following properties: 
\begin{enumerate}
    \item It should reduce the amount of information to be transmitted so that $\xv_\mathrm{deg}$ can be recovered at the receiver with minimal reconstruction error, denoted by $\hat{\xv}_\mathrm{deg}$.
    \item It should preserve perceptual invariances of the image so that $\hat{\xv}_\mathrm{deg}$ can be restored by $R_{\psiv}$ yielding a consistent and perceptually high quality image $\hat{\xv}$.
\end{enumerate}

When $\Am = \mathbf{I}$, our scheme for autoencoding stage is equivalent to standard \gls{DeepJSCC}. However, note that, when the channel \gls{SNR} and bandwidth ratio $\rho$ are low, the receiver cannot always recover the desired $\xv_\mathrm{deg}$, and additional reconstruction error is introduced, following an unknown distribution. Our goal by introducing $\Am$ is to obtain a known degradation operator  to remove the additional noise and to allow good restoration. For instance, a lower resolution image can be transmitted more reliably on the same channel w.r.t. its higher resolution version, yet, it requires further restoration at the receiver to recover its high resolution version with better perceptual quality.



\subsection{Restoration of the Degraded Image}
\label{sec:restoration}

Here, we describe the restoration stage that reverts the known degradation modelled by $\Am$ and improves its perceptual quality while remaining as consistent with $\hat{\xv}_\mathrm{deg}$ as possible.

We denote the pseudo-inverse of $\Am$ via $\Am^\dagger$, which satisfies $\Am \Am^\dagger \Am \equiv \Am$, and can be solved in matrix form using \gls{SVD}. For instance, $\Am^\dagger$ corresponds to upsampling matrix for the downsampling-based degradation operator $\Am$. Given degraded image $\xv_\mathrm{deg}$, \gls{IR} aims to yield $\hat{\xv}$ that does not introduce significant distortion while increasing the perceptual quality of the image~\cite{wang2022zero}. For the former, we introduce a consistency constraint that requires $\Am \hat{\xv} \equiv \xv_\mathrm{deg}$, whereas the perceptual quality requires $\hat{\xv} \sim q(\xv)$, where $q(\xv)$ represents the \gls{GT} image distribution. 

We employ \gls{DDPM}, denoted by $Z_{\psiv}$, which is trained through a progressive denoising task. \gls{DDPM} consists of T-step forward and backward processes. The forward process gradually introduces random noise into the data, whereas the reverse process generates desired data samples from noise. Let $\boldsymbol{\epsilon}_t$ denote the noise in $\xv_t$ at timestep $t$. \gls{DDPM} utilizes a \gls{DNN}, $Z_{\psiv}$, to predict the noise, $\boldsymbol{\epsilon}_t$, i.e., $\boldsymbol{\epsilon}_t=Z_{\boldsymbol{\psiv}}(\mathbf{x}_{t},t)$ is the estimation of $\boldsymbol{\epsilon}_t$ at time-step $t$. We refer the reader to~\cite{wang2022zero,dhariwal2021diffusion} for further details on \glspl{DDPM}. Although \Glspl{DDPM} are powerful unconditional image generators, generating consistent images is challenging~\cite{dhariwal2021diffusion,wang2022zero}. We employ zero-shot image restoration method \gls{DDNM}, which utilizes and guides the \gls{DDPM} model $Z_{\psiv}$ for the restoration task to improve quality while preserving consistency. Note that since \gls{DDNM} utilizes a pretrained \gls{DDPM}, it does not require specific training for restoration task~\cite{wang2022zero}.

Any sample $\xv$ can be decomposed into $\xv \equiv \Am^\dagger \Am \xv + (\mathbf{I} - \Am^\dagger \Am) \xv$, called range-null space decomposition, where the two terms correspond to range and null space parts of the image, respectively. We denote the estimated $\xv_0$ at diffusion time-step $t$ as $\xv_{0|t}$. For a degraded image $\xv_\mathrm{deg}$, we can construct a solution for $\hat{\xv}$ that satisfies the consistency constraint:
\begin{align*}
\hat{\xv} = \Am^\dagger \xv_\mathrm{deg} + (\mathbf{I} - \Am^\dagger \Am) \xv_{0|t},
\end{align*}
where $\xv_{0|t}$ is iteratively refined via the procedure shown in~\Cref{fig:restoration,alg:restoration}, where $\Bar{\alpha}$ is a hyperparameter determining the noise schedule and $p$ is the forward diffusion process distribution (see~\cite{wang2022zero} for more details). Since modifying $\xv_{0|t}$ does not affect the consistency constraint, the restoration procedure does not break the consistency of the transmitted image. Therefore, we refine the image to increase its perceptual quality while keeping it consistent to the version reconstructed by the DeepJSCC decoder with a known degradation.
Finally, we unflatten $\hat{\xv}$ back to dimensions $C_\mathrm{in} \times W \times H$.

    \begin{algorithm}[tbp!]
    \caption{Image restoration procedure $R_{\psiv}$.}
    \label{alg:restoration}
    \begin{algorithmic}[1]
        \State $\mathbf{x}_{T}\sim\mathcal{N}(\mathbf{0},\mathbf{I})$ \Comment{Initialize from pure noise}
        \For{$t = T, ...,  1$}
            \State $\mathbf{x}_{0|t} = \frac{1}{\sqrt{\Bar{\alpha}}_{t}}	\left( \mathbf{x}_{t} - Z_{\psiv}(\mathbf{x}_{t},t)\sqrt{1-\Bar{\alpha}_{t}} \right)$
             \Comment{Denoising step}
            \State $ \hat{\mathbf{x}}_{0|t} = \mathbf{A}^{\dagger} \hat{\xv}_\mathrm{deg} + (\mathbf{I} - \mathbf{A}^{\dagger}\mathbf{A})\mathbf{x}_{0|t}$ \Comment{Refine null-space}
            \State $\mathbf{x}_{t-1}\sim p(\mathbf{x}_{t-1}|\mathbf{x}_{t},\hat{\mathbf{x}}_{0|t})$ \Comment{Sample from reverse process}
        \EndFor
        \State \textbf{return} $\mathbf{x}_{0}$
    \end{algorithmic}
    \end{algorithm}

\subsection{Autoencoder Network Architecture and Training}

 \begin{figure*}[tbp!]
     \centering
     \includegraphics[width=0.85\linewidth]{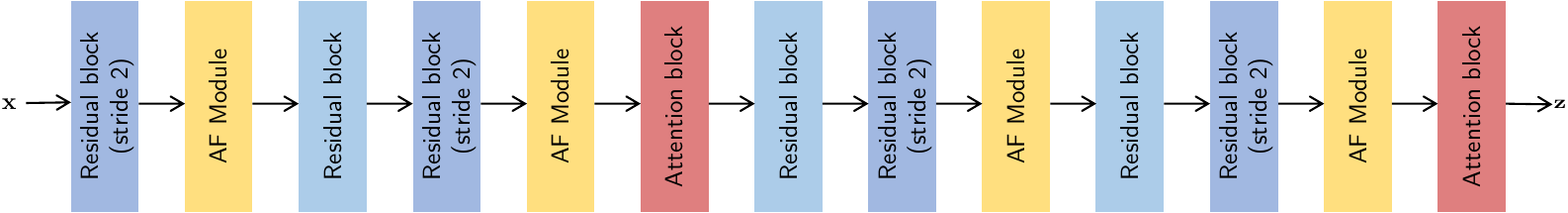}
     \vfill
     \vspace{10pt}
     \includegraphics[width=0.9\linewidth]{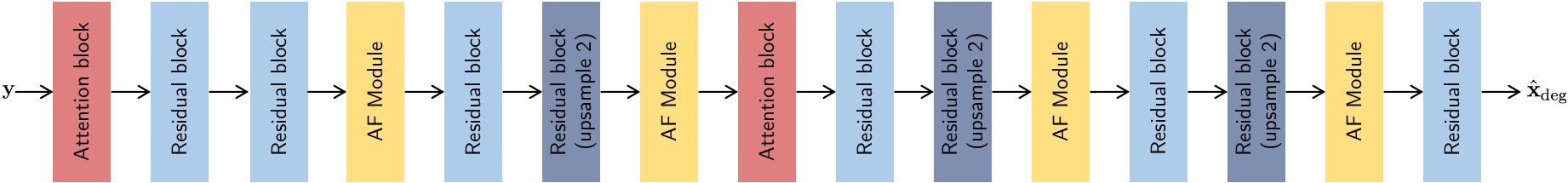}
     \caption{\selim{The encoder (top) and decoder (bottom) architectures of the employed DeepJSCC scheme.}}
     \label{fig:degdeepjscc}
 \end{figure*}

\selim{\Cref{fig:degdeepjscc} shows the details of the employed \gls{CNN}-based autoencoder architecture for \gls{DeepJSCC}. For fair comparison, we employ the same \gls{NN} architecture with~\cite{erdemir2022generative} as our autoencoder, except that we replace the first residual upsample block of the decoder with a residual block to produce the degraded image $\hat{\xv}_\mathrm{deg}$.} 
\glspl{CNN} allow extracting high-level features by exploiting spatial structures within high dimensional inputs such as images, and has been known to perform well for various vision-related tasks~\cite{bourtsoulatze2019deep}.
We employ an architecture that is similar to~\cite{tung2022deepjscc} that has nearly symmetric encoder and decoder.
It also has residual connections and a computationally efficient attention mechanism~\cite{cheng2020learned}. We utilize \gls{AF} module to prevent significant performance degradation for different \glspl{SNR}~\cite{xu2021wireless}.
We randomly sample \gls{SNR} values during training since \gls{AF} module requires \gls{SNR} as an input.

Notice that, unlike the current \gls{DeepJSCC} architectures~\cite{bourtsoulatze2019deep, tung2022deepjscc}, our \gls{CNN}-based autoencoder reconstructs the degraded image $\hat{\xv}_\mathrm{deg}$, which is later restored by $R_\mathbf{\Psi}$. We denote degradation matrix $\Am$ to be the average pooling matrix (with downsampling factor of \num{2} on both dimensions), so we have $2 W^{'} = W$, $2 H^{'} = H$ and $C_{\mathrm{in}}=C^{'}_{\mathrm{in}}$. The reason for this choice is that we can reliably revert this degradation and obtain a high quality image via restoration as described in~\Cref{sec:restoration}.

We first train the encoder and decoder parameters $\Thetav$ and $\Phiv$ to minimize the \gls{MSE} loss:
\begin{align*}
    \Lc \LB \xv, \hat{\xv}_\mathrm{deg} \RB = \frac{1}{C_\mathrm{in} W H} \norm{\Am \xv - \hat{\xv}_\mathrm{deg}}_2^2,
\end{align*}
which aids our DeepJSCC-Degraded model to transmit degraded instances with minimal distortion w.r.t. the known degradation modelled by $\Am$.

\section{Numerical Results}
\label{sec:numerical_results}
\selim{In this section, we present our experimental setup and demonstrate the performance gains of our method.}

\subsection{Experimental Setup}

We evaluate our method on \num{512}x\num{512} CelebA-HQ dataset, which contains \num{30000} high resolution images~\cite{karras2017progressive}. We split the dataset using the same procedure as in~\cite{erdemir2022generative}, i.e., split as $8:1:1$ for training, validation, and testing, respectively. We use the \gls{PSNR} and \gls{LPIPS} metrics to evaluate the quality of the reconstructed images. \gls{LPIPS} is a perception metric ~\cite{zhang2018unreasonable}, which computes the similarity between the activations of two image patches for a pre-defined neural network, such as VGG or AlexNet. Note that, a lower LPIPS score is better since it means that image patches are perceptually more similar.

\selim{\subsection{Implementation Details}}

We have conducted the experiments using Pytorch~\cite{paszke2019pytorch}. We use the same hyperparameters and the same architecture for all the methods. We set the learning rate to \num{1e-4}, number of filters in middle layers to \num{128}, batch size to \num{64} and the power constraint to $\Pavg=1$. We use Adam optimizer~\cite{kingma2014adam}. We continue training until no improvement is achieved for consecutive \num{10} epochs. During training and validation, we run the model using different \gls{SNR} values for each instance, uniformly chosen from $\LSB -5, 5 \RSB$ \si{\decibel}. We test and report the results for each \gls{SNR} value using the same model thanks to the \gls{AF} module. We shuffle the training pairs or instances randomly before each epoch.

In the restoration step, we employ \num{512}$\times$\num{512} unconditional Imagenet \gls{DDPM}, which is fine-tuned for \num{8100} steps from OpenAI's class-conditional \gls{DDPM}~\cite{dhariwal2021diffusion}. We set the number of diffusion timesteps to $T=1000$ with linear schedule. We also utilize the time-travel trick in~\cite{wang2022zero} with \num{100} sampling timesteps.

\subsection{Results}
\newcommand{\comparisonfig}[5]{
    \begin{tikzpicture}[scale=#5] 
    \begin{axis}[
        title={#2},
        xlabel={$\mathrm{SNR}_\mathrm{test} (\si{\decibel})$},
        ylabel={#4},
        error bars/y dir=both,
        error bars/y explicit,
        cycle list/Set2-3,
        cycle multiindex* list={
                mark list\nextlist
                Set2-3\nextlist{solid,solid,solid,dashed,dashed,dashed}\nextlist
        },
        legend style={font=\footnotesize},
        legend to name=comparisons_legend,
        legend columns=3,
    ]
    \foreach \cout / \bw / \bww in {8/192/0.0052,2/768/0.0013} {
        \addplot table[x=snr, y=#3, y error=#3_std, col sep=comma]{results/256_mse_\bww_0.0.csv};
        \addplot table[x=snr, y=#3, y error=#3_std, col sep=comma]{results/#1_separate_cout\cout.csv};
        \addplot table[x=snr, y=#3, y error=#3_std, col sep=comma]{results/#1_deepjscc_cout\cout.csv};
    }
    \legend{Ours ($\rho=0.0052$),Inverse-JSCC ($\rho=0.0052$)~\cite{erdemir2022generative},DeepJSCC ($\rho=0.0052$)~\cite{bourtsoulatze2019deep},Ours ($\rho=0.0013$),Inverse-JSCC ($\rho=0.0013$)~\cite{erdemir2022generative},DeepJSCC ($\rho=0.0013$)~\cite{bourtsoulatze2019deep}}
    \end{axis}
    \end{tikzpicture}
}

\begin{figure}[!t]
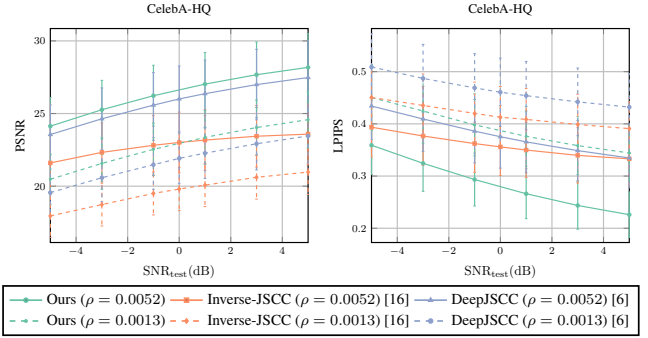

\centering
    \comparisonfig{celeba}{CelebA-HQ}{psnr}{PSNR}{0.5}
    \comparisonfig{celeba}{CelebA-HQ}{lpips}{LPIPS}{0.5}
\resizebox{\columnwidth}{!}{
    \ref*{comparisons_legend}
}
        \caption{\gls{PSNR} and \gls{LPIPS} comparison of our method with the baselines for $\rho \in \{0.0013, 0.0052\}$ over different \glspl{SNR}.}
    \label{fig:comparisons}
\end{figure}

In this section, we show the superiority of our method over standard \gls{DeepJSCC} and \gls{GAN}-based \gls{DeepJSCC}~\cite{erdemir2022generative}, named Inverse-JSCC. Since previous studies have already shown that \gls{DeepJSCC} outperforms classical separation-based methods, we do not consider them as baselines~\cite{bourtsoulatze2019deep}.

\begin{figure}
    \centering
    \begin{tabular}[b]{ccc}
   \footnotesize{Original Image} & \footnotesize{DeepJSCC} & \footnotesize{DeepJSCC-Diff (Ours)} \\
    \begin{subfigure}
        \centering    \includegraphics[width=0.28\linewidth]{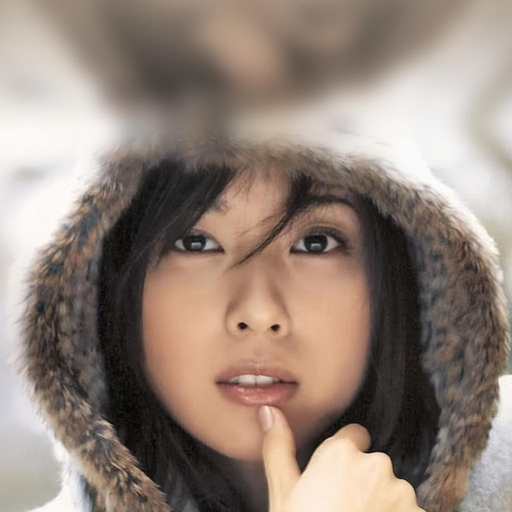}
    \end{subfigure}&
    \begin{subfigure}
        \centering    \includegraphics[width=0.28\linewidth]{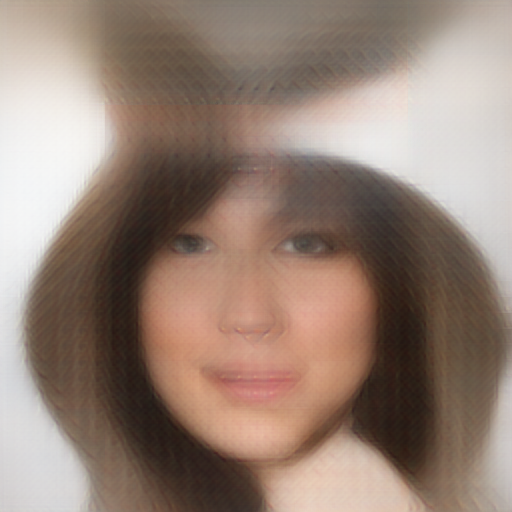}
    \end{subfigure}&
    \begin{subfigure}
        \centering    \includegraphics[width=0.28\linewidth]{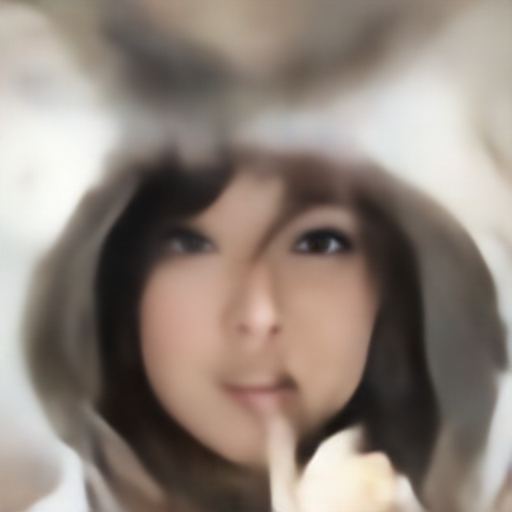}
    \end{subfigure}\\
    \begin{subfigure}
        \centering    \includegraphics[width=0.28\linewidth]{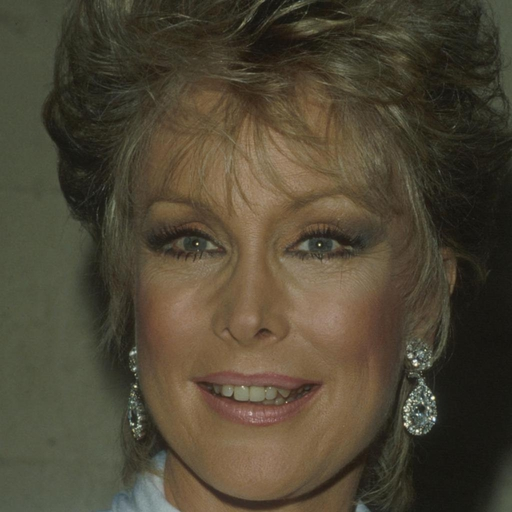}
    \end{subfigure}&
    \begin{subfigure}
        \centering    \includegraphics[width=0.28\linewidth]{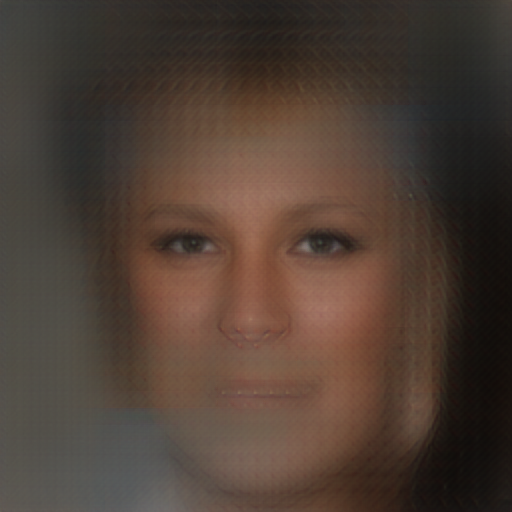}
    \end{subfigure}&
    \begin{subfigure}
        \centering    \includegraphics[width=0.28\linewidth]{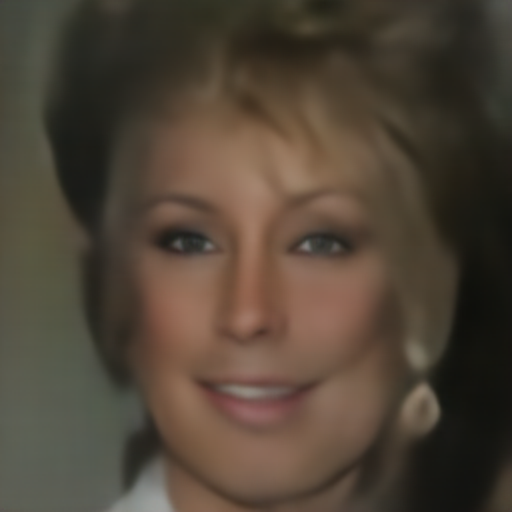}
    \end{subfigure}\\
    \end{tabular}
    \caption{Qualitative comparison of the reconstructed images from CelebA-HQ dataset for $\rho=0.0013$ and $\mathrm{SNR}_{\mathrm{test}}=3$ \si{\decibel}.}
    \label{fig:reconstructed_images}
\end{figure}

\Cref{fig:comparisons} shows the comparisons in terms of \gls{PSNR} and \gls{LPIPS} metrics, respectively. We highlight that we consider an extremely challenging communication scenario with a very low bandwidth ratio. Our method clearly improves w.r.t. both DeepJSCC and Inverse-JSCC at all evaluated values of $\mathrm{SNR}_\mathrm{test} \in \LSB -5 \,, 5 \, \RSB$ \si{\decibel} and $\rho \in \{0.0013,0.0052\}$. \Cref{fig:reconstructed_images} shows an example set of reconstructed images for qualitative comparison. It is clear that the proposed method generates more realistic images, but surprisingly, it also achieves a much lower distortion at all channel conditions.  
We note that our method improves upon DeepJSCC even in terms of the distortion metric, \gls{PSNR}. This shows that it is beneficial for the DeepJSCC encoder/decoder to target a lower resolution version of the image, which can then be improved at the receiver using the diffusion model.

\section{Conclusion}
\label{sec:conclusion}

We have introduced a novel generative communication scheme for image transmission over noisy wireless channels, which promotes realness and consistency through controlled degraded image transmission followed by restoration. The introduced method employs \gls{DDPM} in addition to a modified \gls{DeepJSCC}, which aims at transmitting a degraded image with the degradation modeled as a known linear transform. We have shown that the proposed scheme outperforms the standard \gls{DeepJSCC} and the state-of-the-art \gls{GAN}-based \gls{DeepJSCC}. While we have considered a specific linear transform for the DeepJSCC encoder/decoder pair in this work, we will consider its optimization in our future work. 


\bibliographystyle{IEEEbib}
\bibliography{main}


\end{document}